\documentstyle{elsart}

\renewcommand{\baselinestretch}{2.0}
\begin{document}

{
\renewcommand{\baselinestretch}{1.0}
\begin{frontmatter}
\title{Non-equilibrium surface diffusion in the O/W(110) system}

\author[Helsinki,Providence]{I. Vattulainen\thanksref{Vattu_grants}}, 
\author[Helsinki,Providence,Jyvaskyla]{J. Merikoski},
\author[Helsinki,Providence,Tampere]{T. Ala-Nissila}, and 
\author[Providence]{S. C. Ying}

\address[Helsinki]{Research Institute for Theoretical Physics, 
	P.O. Box 9 (Siltavuorenpenger 20 C), FIN--00014 
	University of Helsinki, Finland}
\address[Providence]{Department of Physics, Box 1843, Brown University, 
	Providence, R.I. 02912, U.S.A.}
\address[Jyvaskyla]{Department of Physics, University of Jyv\"askyl\"a, 
	P.O. Box 35, FIN--40351 Jyv\"askyl\"a, Finland}
\address[Tampere]{Laboratory of Physics, 
	Tampere University of Technology, P.O. Box 692, 
	FIN--33101 Tampere, Finland}
\thanks[Vattu_grants]{Corresponding author. E-mail: Ilpo.Vattulainen@csc.fi.}

\begin{keyword}
Surface diffusion, Growth, Computer simulations, Tungsten, Oxygen
\end{keyword}

\begin{abstract}
In this Letter, we present results of an extensive Monte Carlo 
study of the O/W(110) system under non-equilibrium conditions. 
We study the mean square displacements and long wavelength 
density fluctuations of adatoms. From these quantities, we define 
effective and time-dependent values for the collective and tracer 
diffusion mobilities. These mobilities reduce to the usual 
diffusion constants when equilibrium is reached. We discuss our 
results in view of existing experimental measurements of effective 
diffusion barriers, and the difficulties associated with 
interpreting non-equilibrium data. 
\end{abstract}

\end{frontmatter}
}

\pagebreak

\section{Introduction}

Surface diffusion plays a significant role in many 
surface phenomena such as epitaxial growth, catalysis, and 
ordering \cite{Applications}. These phenomena are interesting 
both from the point of view of basic research and
applied science. Many new experimental techniques 
\cite{Bot96,Swa96,Tri95} have been developed for the 
measurement of surface diffusion  either under equilibrium 
conditions or slight deviations from equilibrium. There is a 
corresponding increase in our theoretical understanding of 
this important quantity. Under near equilibrium conditions, 
the theoretical description of surface diffusion is well 
established based on the linear response theory by Kubo 
\cite{Kub66}. Equivalently, the problem can be described 
by the diffusion equation governing the evolution of the 
density profile \cite{Gom90}. However, in many non-equilibrium 
situations such as surface growth, there is no unique way of 
defining a diffusion constant $D$. Yet the diffusive motion 
of the adatoms clearly plays an important role in such
processes. In this work, we will introduce the concepts of 
tracer and collective {\em mobilities} which characterize 
the motion of adatoms under non-equilibrium situations. These
quantities are defined as generalizations of the usual
diffusion constants to which they reduce in the appropriate
limits. To study the non-equilibrium mobilities, 
we have undertaken an extensive Monte Carlo study of the 
O/W(110) system using a lattice-gas model. We consider the 
non-equilibrium ordering dynamics of this system after a 
quench from a totally random initial state to a temperature 
characterizing an ordered state. We then divide the time 
scales into slices according to the decay of the excess 
energy. In each time slice, we introduce a definition for 
tracer and collective mobility. These mobilities are then 
fitted to an Arrhenius form to extract the effective 
diffusion barriers that are shown to be strongly time-dependent 
during the ordering process. For comparison, the equilibrium 
properties have also been determined. We discuss our results 
in light of the  experimental data by Tringides {\it et al.} 
\cite{Tri87,Lag88,Wu89,TriReview}, and consider the 
difficulties associated with interpreting non-equilibrium 
measurements.

\section{Model and methods}

The phase diagram of the O/W(110) system is fairly well determined 
through a series of experimental studies \cite{Wu89,Chi78,Wan78}. 
Its main features have also been obtained from theoretical 
calculations with a lattice-gas Hamiltonian including pair 
interactions up to fifth neighbors and triplet interactions 
\cite{Sah88}. We use this lattice-gas Hamiltonian for our present 
study of non-equilibrium properties of this system through Monte 
Carlo (MC) simulations. For this work, we focus at the coverage 
$\theta = 0.45$. At this coverage, the experimental value for the 
critical temperature of the order-disorder transition is 
$T_c^{\mbox{\scriptsize exp}} \approx 710$ K \cite{Wan78}. 
Above $T_c$, the adsorbate is in a disordered phase. Below $T_c$,
the system is characterized by an ordered $p(2\times 1)$ or equivalently  
the $p(1\times 2)$ phase. In this work, we study the ordering dynamics 
of the system by starting from an initial state, which is completely 
random such as that obtained from random deposition at low temperatures. 
The system is then allowed to evolve towards the equilibrium ordered 
phase at various temperatures ranging from 0.655 $T_c$ to 0.893 $T_c$. 
This procedure corresponds to instantaneous heating 
in an actual experimental situation.

To perform our MC simulations, we chose the transition
dynamics algorithm (TDA) \cite{Ala92}. Within TDA the transition rate 
$w_{i,f}$, from an initial state $i$ with energy $E_i$ to a final 
state $f$ with energy $E_f$, is decomposed into two steps by introducing 
an intermediate state $I$ with energy $E_I = (E_i + E_f)/2 + \Delta$, 
where the quantity $\Delta$ characterizes the activation barrier in the 
zero coverage limit due to the substrate-adatom interaction. The rate 
$w_{i,f}$ is then the product of the two rates $w_{i,I}$ and $w_{I,f}$, 
which are taken to be of the standard Metropolis form \cite{Bin84}. 
The TDA describes the classical diffusion barrier more 
realistically than other transition rate algorithms \cite{Ala92,Bin84}, 
in which the effect of the saddle point of the adiabatic surface 
potential is not taken into account.

For the present study, we chose $\Delta= 0.0437$ eV. This value is 
believed to be much lower than the true value which should be closer to 
the experimentally observed barrier of 0.5 to 0.6 eV \cite{Wu89,Che79} 
in the disordered phase. Our choice is necessitated by the need to 
speed up the jump rate in the numerical simulations at low temperatures. 
We have done some tests and found that the effect of $\Delta$ and 
the adatom-adatom contributions to the diffusion barrier are approximately 
additive. Therefore the barriers calculated here should be increased
by about 0.5 to 0.6 eV for comparison with the experimentally observed 
values.

To analyse the ordering dynamics in terms of quasi-equilibrium concepts, 
we have to divide the total ordering period into different time regimes. 
These slices of time obviously have to be different at different 
temperatures because of the change in the rate of ordering. For this 
purpose, we first calculate the time-dependence of the excess energy 
of the system, namely $E_x(T,t) = E(T,t) - E(T,\infty)$, after an 
instantaneous quench at time $t=0$ from a completely random state 
to a temperature $T < T_c$. Here $E(T,t)$ is the energy of the system 
at temperature $T$ and time $t$. We then introduce the normalized 
excess energy $F(T,t) = E_x(T,t)/E_x(T,0)$, which has the maximum 
$F(T,0) = 1$. The equivalent time regimes at different temperatures are 
chosen as intervals between times $t_n(T)$ which satisfy 
$F(T,t_n(T)) = \exp(-n)$, where the integer $n\geq 0$.
Typically, we have used five such time slices here.

Now we come to the definition of the non-equilibrium mobilities. 
In the case of tracer mobility, we consider the quantity 
\begin{equation}
\xi_{\alpha\alpha}^{(n)}(\delta t) = \frac{1}{4N} \sum_{k=1}^{N}
        \langle | 
	 R_{\alpha}^{k}(t_n(T) + \delta t) - R_{\alpha}^{k}(t_n(T))
                |^2 \rangle, 
        \label{Eq:neqtd}
\end{equation}
where $\alpha = x,y$ and the sum is over $N$ particles to improve 
statistics. The two independent spatial components ($x$, $y$) of the 
position vector $\vec{R}^{k}(t)$ for a particle $k$ at time $t$ are 
denoted by $R_{\alpha}^{k}(t)$, i.e. 
$\vec{R}^{k}(t) = (R_{x}^{k}(t),R_{y}^{k}(t))$, 
and $\langle \,\,\,\, \rangle$ is used to denote configuration 
averaging. Within each slice of time, the time difference $\delta t$ 
obeys $0 \leq \delta t < t_{n+1}(T) - t_n(T)$. The tracer mobility 
$D_{T,\alpha\alpha}^{(n)}$ is then defined as the effective slope 
of $\xi_{\alpha\alpha}^{(n)}$ within the given time regime $n$. In
equilibrium and for $\delta t \rightarrow \infty$, Eq. (1)
is just the tracer diffusion constant $D_T$ times $\delta t$. 
If the system approaches equilibrium as a function of time, 
$\xi^{(n)}/\delta t$ tends towards $D_T$, too.
  
To study collective mobility, we introduce the time-dependent 
density-fluctuation autocorrelation function 
\begin{equation}
S^{(n)} (\vec{R},\vec{R}',\delta t) = \langle 
	\delta\rho (\vec{R},t_n(T) + \delta t) \delta\rho(\vec{R}',t_n(T)) 
	\rangle, 
\end{equation}
where $\vec{R}$ and $\vec{R}'$ denote position vectors on a lattice, 
and the density fluctuations are given by 
$\delta\rho(\vec{R},t) = \rho(\vec{R},t) - \langle\rho(\vec{R},t)\rangle$. 
Again, $0 \leq \delta t < t_{n+1}(T) - t_n(T)$ within each time regime $n$. 
In the limit of long times (large $n$ and $\delta t$) and in the 
hydrodynamic regime, i.e. in equilibrium, the Fourier transform 
of this correlation function decays as $S(\vec{k},\delta t) = 
S(\vec{k},0) \exp( - \vec{k} \cdot D \cdot \vec{k} \, \delta t )$. 
In the non-equilibrium situation, we consider 
$\log S^{(n)}(\vec{k},\delta t)$ over the given regime $n$ and 
define its effective slope divided by  $k^2$ as the collective 
mobility $D_{C,\alpha\alpha}^{(n)}$. In practice this method was 
carried out with separate sine and cosine transforms of the density 
fluctuations, following the approach in Ref. \cite{Mak88}. A more 
detailed presentation of the various methods used will be given 
elsewhere \cite{Vat96}.

The MC calculations were carried out in a $M\times M$ lattice, 
the system size being $M=30$. Although it is rather small, it is 
large enough when one is not close to the critical region. 
The number of independent samples varied between 
1000 and 10000.

\section{Results and Discussion}

We want to examine first the validity of extracting the mobilities 
from the quantities $\xi_{\alpha\alpha}^{(n)}(\delta t)$ and 
$\log S^{(n)}(\vec{k},\delta t)$ 
by assuming their linear dependence on $\delta t$ within each time slice.
A typical example of the behavior of $\xi_{xx}^{(n)}(\delta t)$ is given 
in Fig. \ref{Figure1}. The deviations from linear behavior are not very 
large at any time, therefore our definition for $D_{T}^{(n)}$ is well 
justified. A similar result is also found for 
$\log S^{(n)}(\vec{k},\delta t)$ used to extract $D_{C}^{(n)}$. 
Furthermore, it is evident from Figs. \ref{Figure2} (a) and 
\ref{Figure2} (b) that within each time slice $n$ the temperature 
dependence of the resulting tracer and collective mobilities are well 
described by an Arrhenius form $D^{(n)} \sim \exp(-\beta E_A)$, with $E_A$ 
denoting the effective activation barrier for diffusion. This allows us 
to determine the time-dependence of $E_A$ during the ordering process. 
The results for $E_A$ based on the tracer and collective mobilities are 
shown in Fig. \ref{Figure3}. Three interesting features emerge. At early 
times, the adsorbate is in a disordered configuration and the adatom-adatom 
interaction contributions to the activation barrier largely cancel out. 
Thus, $E_A$ is approximately just the intrinsic barrier $\Delta$. 
At intermediate times, $E_A$ increases rapidly, with the barrier for the 
tracer mobility always larger than the corresponding one for the 
collective mobility. Finally, at long times $E_A$ approaches the 
equilibrium value $E_A^{\mbox{\scriptsize eq}} = 0.297 \pm 0.008$ eV 
given by the dashed line. We note that, in the equilibrium, $E_A$'s 
for tracer and collective diffusion are indeed equal within error limits.

We now discuss the relevance of our results to some 
experiments on the O/W(110) system by Tringides {\it et al.} 
\cite{Tri87,Lag88,Wu89,TriReview}. These authors studied 
the ordering dynamics of the $p(2\times 1)$ phase at 
$\theta \approx 0.5$ following an up-quench in temperature 
as in our simulation work. They found that the time-dependent 
average domain size $L(t)$ followed the growth law 
$L(t) = A(T) t^{\phi}$ with $\phi \approx 0.28$, where 
$\phi$ is the kinetic growth exponent. Using dimensionality 
arguments, they argued that the prefactor $A(T)$ can be 
related to an effective diffusion constant $D$ through the 
relation $A(T) \propto D^{\phi}$. Thus, the measured temperature 
dependence of $A(T)$ allows the determination of an effective 
diffusion barrier $\tilde{E}_A$. For this they obtained
\cite{Tri87,Wu89} the value of
$\tilde{E}_A^{\mbox{\scriptsize exp}} = 0.61 \pm 0.11$ eV
in contrast to the equilibrium measurements of Gomer {\it et al.}
\cite{Che79,Tri85} who obtained
$E_A^{\mbox{\scriptsize exp,eq}} = 1.0 \pm 0.05$ eV
at $\theta \approx 0.56$ in the ordered phase.
We performed a similar analysis for the ordering 
process in our simulations. We used $E_x^{-1}(T,t)$ 
as a measure for $L(t)$ \cite{Fog88} and found that the growth 
law is valid over a certain period of time only; namely from 
the end of an initial transient period to the point when the 
finite size or pinning effects come into play. In our case,
we obtain $\phi \approx 0.5$ in the regime corresponding to 
the time slices $n=2,3$. This is illustrated in Fig. \ref{Figure4}. 
We note that at large times, the growth exponent $\phi$ 
is not well defined due to finite-size effects that are 
rather pronounced after the regime $n = 3$. The value of the 
effective barrier $\tilde{E}_A $ extracted from the relation 
$A(T) \propto D^{\phi}$ is $0.192 \pm 0.021$ eV. Our time-dependent 
results for $E_A$ based on the mobilities $D_C^{(n)}$ for these 
intermediate time regimes range between 0.88 $\tilde{E}_A $ 
and 1.1 $\tilde{E}_A $. Thus, we conclude that the procedure of 
using the dimensional analysis to extract an effective barrier 
from the prefactor in the power law growth is indeed valid, and 
it represents the barrier for the dominant diffusion process in 
that time regime. Concerning the absolute values of the diffusion 
barriers, we have to add about 0.6 eV to them as discussed earlier 
to account for our choice of $\Delta$. This brings our results for 
the effective barrier in the power law growth regime to 0.8 eV 
and the final equilibrium barrier to 0.9 eV. The equilibrium 
value is in agreement with the experimental value. Also we see 
the same that the adatom-adatom contribution is most dominant in the
equilibrium ordered phase and the effective barrier measured during 
the ordering processes has a smaller value than the equilibrium 
barrier \cite{Comparison,Cov068}. The discrepancies between 
our calculated value for $\tilde{E}_A$ and the measured value 
are not too surprising, since $\phi \approx 0.28$ is considerably 
lower than the value $\phi \approx 0.5$ observed in our 
simulations. For the one-component $p(2\times 1)$ phase whose 
order parameter is not conserved, the theoretically expected 
value for $\phi$ is 1/2 \cite{Fog88}. 
Experimentally this value has been observed in the O/W(112) 
system \cite{Zuo88}, for example. The experimental 
data for the O/W(110) system 
in Refs. \cite{Tri87,Wu89} probably belongs to the 
early time rather than the intermediate time regime, since the 
maximum mean diameter of the growing domains during the ordering 
experiments is smaller than six lattice spacings \cite{Tri87,Wu89}. 
The $\tilde{E}_A^{\mbox{\scriptsize exp}}$ 
they measure though represents the effective 
diffusion barrier for a different configuration than the one 
in the intermediate time regime. The exact cause for the failure 
of the experiment to reach the $t^{1/2}$ growth regime is still 
unclear \cite{TriReview}.

We believe that in other non-equilibrium methods such as profile 
evolution techniques, useful information for an effective mobility 
is also best extracted from an intermediate time regime. Otherwise, 
the initial behavior with large deviations from equilibrium behavior 
would yield results difficult to interpret. Unlike the theoretical 
simulation studies where we can define the intermediate regimes 
rather precisely, the practical difficulties in actual non-equilibrium
measurements would be to identify the proper intermediate time regimes. 
In the domain ordering dynamics, we have seen that the power law growth 
regime can be identified for this purpose. For other non-equilibrium
situations, it is not clear how to establish experimentally similar 
criteria.

\bigskip\bigskip
{\bf Acknowledgements}

I. V. thanks the Neste Co. Foundation, the Jenny and Antti Wihuri 
Foundation, and the Finnish Academy of Sciences for support. 
J. M. is supported by the Academy of Finland and Emil Aaltonen 
Foundation. This research has also been partially supported 
by a grant from the office of Naval Research (S. C. Y. and 
J. M.). Finally, computing resources of the University of 
Helsinki and the University of Jyv\"askyl\"a are gratefully 
acknowledged.

\begin{figure}[htb]
	\vspace{30mm}
	\caption{Results for $\xi_{xx}^{(n)}(\delta t)$ with five values 
	of $n$, after a quench at $t=0$ from a completely random state 
	to $T = 0.714$ $T_c$. Time difference $\delta t$ is given in 
	Monte Carlo steps (mcs). In the inner figure, a part of the 
	small time behavior has been magnified. The results for 
	$\xi_{yy}^{(n)}$ (not shown here) are similar. \label{Figure1}}
\end{figure}

\begin{figure}[htb]
	\vspace{30mm}
	\caption{(a) Results for the tracer mobilities $D_{T,xx}^{(n)}$ as 
	an Arrhenius plot. In addition to the equilibrium case that is given 
	for the purpose of comparison (stars and dotted line), results up 
	to $n=4$ are presented. The results for $D_{T,yy}^{(n)}$ (not shown 
	here) are similar. (b) A similar plot in the case of collective 
	mobility $D_{C,xx}^{(n)}$. Due to separate sine and cosine 
	transforms of the density fluctuations, two values for each pair 
	of $T$ and $n$ are given. The results for $D_{C,yy}^{(n)}$ (not 
	shown here) are alike. \label{Figure2}}
\end{figure}

\begin{figure}[htb]
	\vspace{30mm}
	\caption{Results for the effective activation energy 
	$E_A$ versus time. Results with five values of $n$ based on the 
	tracer and collective mobility are shown with full squares and 
	open circles, respectively. Error bars are smaller then the size 
	of the symbol for tracer and approximately of the same size for 
	collective mobility. The equilibrium limit 
	$\tilde{E}_A / E_A^{\mbox{\scriptsize eq}} = 1$ with 
	$E_A^{\mbox{\scriptsize eq}} = 0.297$ eV is given by the 
	dashed line. The fit for $\protect\tilde{E}_A$ based on the 
	$A(T)$ data is given in the inset. (See text for details.)
	\label{Figure3}}
\end{figure}

\begin{figure}[htb]
	\vspace{30mm}
	\caption{A typical example of the time evolution of $E_{x}^{-1}$
	data at $T = 0.655$ $T_c$ as a log-log plot. Time is given in Monte
	Carlo steps (mcs), and the system size used was $120\times 120$.
	A reference curve with $\phi = 1/2$ (dotted line) over the regimes 
	$n=2,3$ is also given. The kinetic
	growth exponent $\phi$ and the $A(T)$ data were evaluated between
	6.9 and 8.3 for $\log t$, yielding 0.514 for $\phi$.
	\label{Figure4}}
\end{figure}


\begin{thebibliography}{100}
\bibitem{Comparison} 
We point out that 
we are also aware of other non-equilibrium measurements 
\protect\cite{But77,Bow80} that gave $E_A > 1$ eV. Due to the 
disordered state during the measurements, these results 
seem rather high compared with Gomer's equilibrium results 
\protect\cite{Che79,Tri85}, and are possibly due to impurities 
\protect\cite{Lag88} or reconstruction \protect\cite{Gom90}.
\bibitem{Cov068} 
We like to mention that at $\theta \approx 0.68$, a similar ordering 
experiment has been carried out by Tringides \cite{Tri90}. In this 
measurement, the quality of the data seems to be better than in 
previous experiments at $\theta \approx 0.5$ \cite{Tri87,Wu89}. 
However, since the final equilibrium phase at the end of the ordering 
process is a coexistence phase of $p(2\times 1)$ and $p(2\times 2)$, 
and the equilibrium barrier in this phase is unknown, we have avoided 
comparison with this data.
\end{thebibliography}
\end{document}